\documentstyle[aps,prl,multicol]{revtex}

\author{Daniel Jonathan\cite{djonat} and Martin B. Plenio\cite{plenio}}
\title{Minimal conditions for local pure-state entanglement manipulation}
\address{Blackett Laboratory, Imperial College, London SW7 2BZ, United Kingdom.}
\date{\today}
\begin{document}
\draft

\maketitle

\begin{abstract}
We find a minimal set of necessary and sufficient conditions for 
the existence of a local procedure that converts a finite pure 
state into one of a set of possible final states. This result provides 
a powerful method for obtaining optimal local 
entanglement manipulation protocols for pure initial states. 
As an example, we determine analytically the optimal distillable entanglement 
for arbitrary finite pure states. We also construct an explicit protocol achieving 
this bound.
\end{abstract}


\begin{multicols}{2}

The existence of non-local correlations, or entanglement, between parts of a
composite quantum system is at the heart of quantum information theory and
its applications \cite{Qinfo}. In recent years, much effort has been
expended on the problem of how to define and quantify the entanglement of a
given state in physically meaningful ways. One
very fruitful approach, first pursued by Bennett and co-workers \cite
{Bennett1,Bennett2,Bennett3}, is to regard entanglement in terms of the {\it %
limitations} that exist to the manipulation of a composite system when each
subsystem is operated on locally. A paradigmatic situation is as follows:
suppose Alice and Bob each possess part of a quantum system, which is
prepared in a state $\rho $. Qualitatively, the existence of entanglement
implies that some transformations of $\rho $ which are in principle possible
cannot be realised if Alice and Bob are only allowed to perform {\it local}
operations on their respective subsystems, and to exchange classical
communication. (Transformations of this type can be referred to 
as `local transformations', or `LQCC' for short). 

A quantitative way of expressing this fact is in terms of so-called {\it %
entanglement monotones }(EMs) \cite{Guifre}. These are functions $%
\varepsilon \left( \rho \right) $ of the quantum state that can, on average,
never increase under LQCC \cite{Monotones}. There
are many known EMs, for example the entanglements of distillation \cite
{Bennett1,Bennett2,Bennett3} and formation \cite{Bennett3,Wootters}, and the
relative entropy of entanglement \cite{Vlatko} (in fact, any reasonable
measure of entanglement must by definition be an entanglement monotone, and
vice-versa). Despite their different physical interpretations, they all
share a common feature:\ a transformation which, on average, increases {\it %
any single }EM cannot be realised locally. In other words, they provide {\it %
necessary }conditions any local transformation $T$ must satisfy.

A natural question that presents itself is then: what are {\it sufficient }%
conditions for $T$ to be local? In other words, we would like to have a set $%
\left\{ \varepsilon _{i}\right\} $ of entanglement monotones such that, if 
the average $\left\langle \varepsilon _{i}\left( T\left[ \rho \right] \right)
\right\rangle \leq \varepsilon _{i}\left( \rho \right) $ for all $i$, then $T
$ is local. Ideally, this set should also be {\it minimal, }in the sense
that these conditions should not be redundant \cite{Guifre2}. An important
result in this direction was recently presented by Nielsen \cite
{Nielsen}, who found sufficient conditions for the locality of
transformations that take one given {\it pure} state to another with 100\%
probability. In the present Letter, we extend Nielsen's theorem to the case
where the transformation need not be deterministic, that is, when $T$ may
lead to several possible final states. We demonstrate that, for
this case, a set of EMs recently introduced by Vidal \cite{Guifre} is in
fact minimal in the sense described above. They therefore provide us with a
powerful universal tool for finding optimal local entanglement manipulation
protocols. We apply it to the problem of {\it entanglement concentration }%
(or {\it purification}), which concerns understanding to which extent
distant parties can extract a maximally entangled state from a non-maximally
entangled one using only LQCC \cite{Bennett1,Bennett2,Bennett3,Gisin}. This
is one of the central problems of quantum information theory, and is of
crucial importance for all applications, such as teleportation \cite{telep},
which require the existence of maximally entangled states between
distant parties. With the help of our generalisation of Nielsen's
theorem, and of results from the well-known simplex optimization method of
linear programming{\it \ }theory \cite{Linprog}, we are able to analytically
determine the {\it optimal} purification protocol for the case where Alice and Bob
share a given pure state $\left| \psi \right\rangle $. By `optimal' we mean
the following:\ assume that Alice and Bob locally manipulate their shared
state until they obtain either a maximally entangled state (of some
dimension), or a completely disentangled one. We determine the strategy that
awards them, on average, with the largest amount of distilled
entanglement, which we find to be 
\begin{equation}
\left\langle E\right\rangle _{\max }=\sum_{j=1}^{N}\left( \alpha _{j}-\alpha
_{j+1}\right) j\ln j, \label{equation1}
\end{equation}
where $\alpha _{1}\geq ...\geq \alpha _{N}$ are the nonzero Schmidt
coefficients of $\left| \psi \right\rangle $.  

It is important to stress that our results pertain to any finite{\it \ }%
shared state. Until now (see also endnote), the problem of finding the best 
purification protocol in the sense above had been completely solved 
(for pure states and some particular mixed states), only in the {\it asymptotic limit},
where Alice and Bob share $N\rightarrow \infty $ identical copies of the
same state \cite{Bennett1,Bennett3}. This limit has fundamental significance
in quantum information and communication theory, for example for deriving
bounds on channel capacities \cite{Bennett3}. Nevertheless, it is clear that
in practise Alice and Bob will always share only a {\it finite}, in general
small,{\it \ }amount of entanglement. Thus, as a number of authors \cite
{Guifre,Guifre2,Nielsen,Lo} have stressed, it is also important to
understand entanglement transformations in this regime, with the asymptotic
results emerging in the suitable limit.

Suppose then that Alice and Bob share a pure state $\left| \psi
\right\rangle $ of a bipartite quantum system, with ordered Schmidt decomposition
$\left| \psi \right\rangle =\sum_{i=1}^{N}\sqrt{\alpha _{i}}\left|
i_{A}\right\rangle \left| i_{B}\right\rangle$ \cite{remark3}.
Vidal \cite{Guifre} has shown that each of the following set of functions of
the $\alpha _{i}$ constitutes an entanglement monotone: 
\begin{equation}
E_{l}\left( \left| \psi \right\rangle \right) =\sum_{i=l}^{N}\alpha _{i},
\;\; 1\leq l\leq N.
\end{equation}

We can use these monotones to describe the following theorem due to Nielsen 
\cite{Nielsen}: let $\left| \eta \right\rangle $ be another pure bipartite
state. Then there exists a local transformation that takes $\left| \psi
\right\rangle $ to $\left| \eta \right\rangle $ with 100\%\ certainty iff $%
E_{l}\left( \left| \eta \right\rangle \right) \leq E_{l}\left( \left| \psi
\right\rangle \right) $, $2\leq l\leq N$. In other words, the $\left\{
E_{l}\right\} $ form a sufficient set of monotones for this kind of
transformation. In fact, since they also uniquely determine the Schmidt
components of $\left| \psi \right\rangle$ and $\left| \eta \right\rangle$, which
completely and minimally characterise such transformations (sec. 5.1 of 
\cite{Guifre2}), it follows that $\left\{ E_{l}\right\} $ is actually a {\it %
minimal }set of EMs in this case.

Quantum mechanics is not, however, concerned only with deterministic
transformations. As long as Alice and Bob do not lose or discard information
about their system, the most general transformation they can apply on $%
\left| \psi \right\rangle $ will produce one of $m$ possible pure states 
$\left| \eta _{i}\right\rangle $, with probability $p_{i}$.{\em \ }We
demonstrate now that Vidal's monotones also provide necessary and sufficient
conditions for these general transformations to be realised locally.

{\bf Theorem 1: }Let 2 distant parties share a pure state $\left| \psi
\right\rangle $ ; let $\left\{ \left| \eta _{j}\right\rangle \right\}
_{j=1}^{m}$ be a set of $m$ other pure bipartite states. Then a
transformation $T$ of $\left| \psi \right\rangle $ that outputs state $%
\left| \eta _{j}\right\rangle $ with{\em \ }probability $p_{j}$\ $\left(
\sum_{j}p_{j}=1\right) $ can be realised using LQCC iff the $N$
entanglement monotones $E_{l}$ do not increase on average, that is, iff 
\begin{equation}
\sum_{j=1}^{m}p_{j}E_{l}\left( \left| \eta _{j}\right\rangle \right) \leq
E_{l}\left( \left| \psi \right\rangle \right) ,1\leq l\leq N.  \label{minimal}
\end{equation}

{\it Proof:\ }Necessity follows from the definition of an entanglement
monotone, and is proven for the $E_{l}$ functions in \cite{Guifre}. To prove
sufficiency, assume eq. $\left( \ref{minimal}\right) $ is satisfied. We will
construct an explicit local strategy that realises the transformation $T.$
First of all, it is clear that we only need to consider the special case
where all target states $\left| \eta _{j}\right\rangle $ have the same
Schmidt basis as $\left| \psi \right\rangle $ (which we can refer to as the
`standard' basis). The general case then follows from the following simple
facts: $\left( i\right) $ any two states with the same Schmidt components
are interconvertible by a local unitary operation, so that to realise $T$
one only needs to generate, with probability $p_{j}$, a state $\left| \eta
_{j}^{\prime }\right\rangle $ with the same Schmidt coefficients as $\left|
\eta _{j}\right\rangle $ in the standard basis. $(ii)$ If two or more target
states $\left| \eta _{j_{1}}\right\rangle ,...,\left| \eta
_{j_{n}}\right\rangle $ have {\it exactly} the same Schmidt components, one
can generate the state $\left| \eta _{j_{1}}^{\prime }\right\rangle $ with
probability $\sum_{k=1}^{n}p_{j_{k}}$, and then `roll a classical die' with
relative probabilities $\frac{p_{j_{i}}}{\sum_{k=1}^{n}p_{j_{k}}}$ to decide
which one of the $\left| \eta _{j_{k}}\right\rangle $ to transform to.

Suppose then that the target states can all be written in the ordered Schmidt form
$\left| \eta _{j}\right\rangle =\sum_{i=1}^{N}\sqrt{\mu _{ji}}\left|
i_{A}\right\rangle \left| i_{B}\right\rangle$
(note that the number of nonzero Schmidt components of $\left| \eta
_{j}\right\rangle $ cannot be greater{\it \ }than $N$ \cite{Lo}).

Let us now define the {\it average target state }$\left| \bar{\eta}%
\right\rangle $ as 
\begin{equation}
\left| \bar{\eta}\right\rangle \equiv \sum_{i=1}^{N}\sqrt{\gamma _{i}}\left|
i_{A}\right\rangle \left| i_{B}\right\rangle, \;\; \gamma
_{i}=\sum_{j=1}^{m}p_{j}\mu _{ji}
\end{equation}
It can be seen that $\gamma _{i}\geq \gamma _{i+1}$, so 
\begin{equation}
E_{l}\left( \left| \bar{\eta}\right\rangle \right)
=\sum_{i=l}^{N}\sum_{j=1}^{m}p_{j}\mu _{ji}=\sum_{j=1}^{m}p_{j}E_{l}\left(
\left| \eta _{j}\right\rangle \right) \leq E_{l}\left( \left| \psi
\right\rangle \right)  \label{average}
\end{equation}
where we have used condition $\left( \ref{minimal}\right) $. We can
therefore apply Nielsen's theorem, which implies that there exists a local
protocol $L$ for {\it deterministically }converting from $\left| \psi
\right\rangle $ to $\left| \bar{\eta}\right\rangle $. Let us now define the
following set of positive operators on Alice's subspace: 
\begin{equation}
A_{j}=\sum_{i=1}^{N}\sqrt{\frac{p_{j}\mu _{ji}}{\gamma _{i}}}\left|
i_{A}\right\rangle \left\langle i_{A}\right| ,1\leq j\leq m
\end{equation}
We can see that, for $1\leq j\leq m:$%
\begin{eqnarray}
A_{j}\otimes {\bf 1}_{B}\left| \bar{\eta}\right\rangle &=&\sum_{i=1}^{N}%
\sqrt{p_{j}\mu _{ji}}\left| i_{A}\right\rangle \left| i_{B}\right\rangle =%
\sqrt{p_{j}}\left| \eta _{j}\right\rangle ,  \label{povm} \\
\sum_{j=1}^{m}A_{j}^{\dag }A_{j} &=&\sum_{i=1}^{N}\left( \frac{%
\sum_{j=1}^{m}p_{j}\mu _{ji}}{\gamma _{i}}\right) \left| i_{A}\right\rangle
\left\langle i_{A}\right| ={\bf P},
\end{eqnarray}
where {\bf P }is the projector $\sum_{i=1}^{N}\left| i_{A}\right\rangle
\left\langle i_{A}\right| $. Together with the complement ${\bf 1}_{A}-{\bf P%
}$, the set $\left\{ A_{j}\right\} _{j=1}^{m}$constitutes therefore a local
POVM\ which, if applied to $\left| \bar{\eta}\right\rangle $, outputs state $%
\left| \eta _{j}\right\rangle $ with probability $p_{j}$. The combination of
this POVM\ with the deterministic protocol $L$ realises the required
transformation $T_{\Box}$

This result can be directly extended to the case where the target
states may be mixed. In this case, eq. $\left( \ref{minimal}%
\right) $ still holds (substituting $\rho _{j}$ for $\left| \eta
_{j}\right\rangle $), as long as we extend the definition of $E_{l}$ using
the `convex roof' rule \cite{Guifre} 
\begin{equation}
E_{l}\left( \rho _{j}\right) =\min_{\rho _{j}=\sum q_{ij}|\eta _{ij}\rangle
\langle \eta _{ij}|}\sum_{j=1}^{m}q_{ij}E_{l}\left( \left| \eta
_{ij}\right\rangle \right)
\end{equation}
where the minimum is taken over all realisations of $\rho _{j}$.

Theorem 1 provides a powerful tool for optimizing local quantum
transformations according to a wide range of criteria. For example,
in ref. \cite{Guifre}, the author seeks to determine the local
transformation that maximizes the probability of converting one given pure
state to another. He obtains an upper bound on this probability from the
existence of the monotones $E_{l}$, and then constructs an explicit protocol
realising the bound. Theorem 1 justifies this result, showing that a similar
strategy will work for {\it any} optimization problem involving an initial pure state;
in other words, the optimum given the constraints expressed in eq. $\left( 
\ref{minimal}\right) $ will always be achievable.

We can immediately apply this result to the problem of optimally
concentrating the entanglement of a finite bipartite pure state. This
situation has already been considered by Lo and Popescu \cite{Lo}, who have
obtained the local protocol that gives the greatest probability of
converting a given pure state $\left| \psi \right\rangle $ to a maximally
entangled state of any {\it given }number of levels. However, it may well be
that Alice and Bob merely wish to concentrate their entanglement, without
regard to what maximally entangled state they end up with. In this case, a
reasonable question to ask is: out of all such local concentration
protocols, which one leads, on average, to the largest amount of shared
distilled entanglement?

The problem may be formally posed as follows: let Alice and Bob share a
single pure state $\left| \psi \right\rangle =\sum_{i=1}^{N}\sqrt{\alpha _{i}%
}\left| i_{A}\right\rangle \left| i_{B}\right\rangle$, whose entanglement 
they wish to concentrate using
LQCC. Following the notation of ref. \cite{Lo}, let us define $\left| \phi
_{j}\right\rangle =\frac{1}{\sqrt{j}}\sum_{i=1}^{j}\left| i_{A}\right\rangle
\left| i_{B}\right\rangle $ as a maximally entangled state of $j$ levels
(note that $\left| \phi _{1}\right\rangle $ is a product state). 
Consider the set of local transformations that generate $\left|
\phi _{j}\right\rangle ,1\leq j\leq N$, with probability $p_{j}$ \cite
{remark2}. If we
choose to measure the amount of entanglement in $\left| \phi
_{j}\right\rangle $ by the von Neumann entropy of $tr_{B}\left| \phi
_{j}\right\rangle \left\langle \phi _{j}\right| $, namely $\ln j$, then the
average amount of distilled entanglement obtained from such a procedure is 
\begin{equation}
\left\langle E\right\rangle =\sum_{j=1}^{N}p_{j}\ln j.  \label{averent}
\end{equation}
Our problem is to maximize this quantity over all probability distributions
for the $p_{i}$ that are consistent with the constraints in eq. $\left( \ref
{minimal}\right) $. Theorem 1 then guarantees the existence of a local
protocol leading to this optimal distribution.

It is easily seen that, for $l \leq j$,
\begin{equation}
E_{l}\left( \left| \phi _{j}\right\rangle \right) =\frac{j-l+1}{j},
\end{equation}
and that it vanishes otherwise. In this case, the constraints in eq. (\ref{minimal})
read 
\begin{equation}
\begin{array}{ll}
\sum_{j=l}^{N}p_{j}\left( \frac{j-l+1}{j}\right) \leq \sum_{j=l}^{N}\alpha
_{j}, & 1\leq l\leq N
\end{array}
.  \label{constraints}
\end{equation}
This is a linear optimization problem with linear inequality constraints, a
kind widely studied in many fields of science and engineering. It can be
solved using the techniques of {\it linear programming }theory, a branch of
applied linear algebra that is familiar to most engineers, though not
so well-known among physicists. We will not attempt to explain the
terminology and results from this theory that are required for our solution;
instead, we refer the reader to textbooks (e.g., \cite{Linprog}).
Our main result is

{\bf Theorem 2: }The optimal entanglement concentration procedure for a
single pure bipartite state $\left| \psi \right\rangle $ with Schmidt
coefficients $\alpha _{1}\geq ...\alpha _{N}>0$ is one that produces a
maximally entangled state $\left| \phi _{j}\right\rangle $ of $j\leq N$
levels with probability $p_{j}^{opt}=j\left( \alpha _{j}-\alpha
_{j+1}\right) $. The corresponding optimal average distilled entanglement is 
$\left\langle E\right\rangle _{\max }=\sum_{j=1}^{N}\left( \alpha
_{j}-\alpha _{j+1}\right) j\ln j.$

{\it Proof: } First, it is easy to check that this probability distribution
satisfies (actually, {\it saturates}) all the inequalities in eq. $\left( 
\ref{constraints}\right) $. In matrix form, we have ${\bf B}\vec{p}=\vec{q}$%
, where $\vec{p},$ $\vec{q}$ are vectors with components $p_{j}=j\left(
\alpha _{j}-\alpha _{j+1}\right) ;q_{l}=\sum_{j=l}^{N}\alpha _{j}$, and $%
{\bf B}$ is an upper triangular $N\times N$ matrix with components $b_{lj}=%
\frac{j+1-l}{j}$ for $j\geq l,$ and $0$ otherwise. In the parlance of linear
programming theory, this is a {\it basic, feasible }solution to the problem,
with all the {\it slack variables} assuming the value zero. We can then
apply the {\it simplex algorithm} to check whether this is the optimal
solution or, if not, to find a better one. A sufficient condition for
optimality (\cite{Linprog}, eqs. (2.36,2.37)) is that the following
inequalities are all satisfied 
\begin{equation}
z_{k}\equiv \sum_{i=1}^{N}c_{i}\beta _{ik}\geq 0, \;\; 1\leq k\leq N
\label{optimality}
\end{equation}
where $c_{i}=\ln i$ is the coefficient of $p_{i}$ in eq. $\left( \ref
{averent}\right) $, and $\beta _{ik}$ are the elements of the inverse of $%
{\bf B.}${\bf \ }It is easy to show that the only nonzero $\beta_{ik}$ are 
\begin{equation}
\begin{array}{lll}
\beta _{k-2,k}=k-2; & \beta _{k-1,k}=-2\left( k-1\right) ; & \beta _{kk}=k.
\end{array}
\end{equation}
The conditions in eq. $\left( \ref{optimality}\right) $ are then trivially
satisfied for $k=1,2$. For $k\geq 3$, we have 
\begin{equation}
z_{k}=\left( k-2\right) \ln \left( k-2\right) +k\ln \left( k\right) -2\left(
k-1\right) \ln \left( k-1\right) .
\end{equation}
The remaining inequalities follow from the convexity of $x\ln x$ for $x>0$ .
The distribution $p_{j}^{opt}=j\left( \alpha _{j}-\alpha _{j+1}\right) $ is
therefore optimal$_{\Box}$

In the remainder of this article, we examine some aspects of Theorem 2.
First of all note that, though Theorem 1 provides an explicit local protocol
realising the optimal probability distribution given above, it is a complicated
one, involving a series of local measurements and subsequent conditioned
local rotations by Alice and Bob. We can, however, also explicitly construct a simpler
optimal protocol, involving only a {\it single} local generalised
measurement (such a simple protocol always exists for {\it any} local
transformation on a bipartite pure state \cite{Lo}). Consider the positive
operators 
\begin{equation}
O_{j}=\sum_{i=1}^{j}\sqrt{\frac{\alpha _{j}-\alpha _{j+1}}{\alpha _{i}}}%
\left| i_{A}\right\rangle \left\langle i_{A}\right| \otimes {\bf 1}_{B}.
\end{equation}
It is easily seen that 
\begin{eqnarray}
O_{j}\left| \psi \right\rangle &=&\sqrt{\alpha _{j}-\alpha _{j+1}}%
\sum_{i=1}^{j}\left| i_{A}\right\rangle \left| i_{B}\right\rangle =\sqrt{%
p_{j}^{opt}}\left| \phi _{j}\right\rangle \\
\sum_{j=1}^{N}O_{j}^{\dag }O_{j} &=&\sum_{i=1}^{N}\frac{\sum_{j=i}^{N}\left(\alpha
_{j}-\alpha _{j+1}\right)}{\alpha _{i}}\left| i_{A}\right\rangle \left\langle
i_{A}\right| ={\bf P}
\end{eqnarray}
where we have interchanged $\sum_{j=1}^{N}\sum_{i=1}^{j}\leftrightarrow
\sum_{i=1}^{N}\sum_{j=i}^{N}$, and where {\bf P }is as in eq. (\ref{povm}). 
The set $\left\{ O_{j},{\bf 1-P}\right\} $ corresponds thus to a
single local POVM measurement that optimally concentrates the entanglement
of state $\left| \psi \right\rangle $.

Although it is optimal, the protocol provided by Theorem 2 is also in general
{\it irreversible}, i.e., it is impossible to recover the original state with
100\% probability. This follows since, in general,
$\left\langle E\right\rangle _{\max } < S $ , where $S$ is the entropy of entanglement
of $\left| \psi \right\rangle$. Note that, since the monotones ${E_l}$ are all 
conserved in this process (the inequalities in eq. (\ref{constraints}) are all 
saturated), this set is {\it not}
sufficient to indicate the reversibility of a local transformation. It can be 
shown, however, that our protocol does becomes reversible in the asymptotic limit
where Alice and Bob share $N\rightarrow\infty$ copies of identical pure states 
(in which case $\left\langle E\right\rangle _{\max } \rightarrow S)$ .
This result, which recovers the one obtained by Bennett et al. \cite
{Bennett1} can be derived from expression (\ref{equation1}) for $\langle E\rangle_{max}$ 
using the saddle point method.
 It can also be checked that, for any finite pure state, our protocol is always 
more efficient than the one suggested in \cite{Bennett1}. This is not surprising, as
their protocol is state independent, while ours is state-dependent.

The solution provided by Theorem 2 has an intuitive appeal: the optimal
protocol for concentrating entanglement is one that first maximizes $p_{N}$,
that is, the likelihood of obtaining the most entangled state possible;
then, given this, it maximizes $p_{N-1}$, and so forth. Although this seems
very reasonable, it is not at all obvious that it should be the case: for
instance, it could have conceivably been more advantageous not to attempt to
obtain $\left| \phi _{N}\right\rangle $, if this choice had sufficiently
increased the likelihood of generating $\left| \phi _{N-1}\right\rangle $
(i.e., enough to increase the final average in eq. $\left( \ref{averent}%
\right) $)$.$ In fact, it can be readily seen that a {\it different }optimal
solution may be obtained if Alice and Bob choose to use a different entanglement 
measure to `weigh' each probability in eq. $\left( \ref{averent}\right) $. As a
simple example: if they use the trivial `indicator' measure that assigns a
value $0$ to a disentangled state, and $1$ to {\it any }entangled state \cite
{Vlatko}, then the optimal solution is the one that maximizes $p_{2}$%
. (This follows from the fact that, for any $j>2$, $\left| \phi
_{j}\right\rangle $ may be locally converted to $\left| \phi
_{2}\right\rangle $ with 100\% efficiency \cite{Lo}). This solution will in general 
{\it not }maximize $p_{N}$ \cite{Lo}, so it differs from the one found in Theorem 2.
Ultimately, the choice of which measure to use (and in the finite-state regime, 
there are many possibilities \cite{Guifre2}) depends on Alice and Bob's particular
needs. Whatever the choice, however, the techniques of Theorems 1 and 2 always 
determine the optimal protocol.

To summarise: we have presented a general method for determining the locality 
of transformations on a given pure bipartite state, based on the nonincrease
of a minimal set of entanglement monotones. We have then used this method to determine
the optimal strategy for locally concentrating the entanglement in such a state. 
We believe that a similar approach will also prove fruitful for more general problems 
involving mixed and/or multiparticle states \cite{remark4}.

We would like to thank P.L. Knight, S. Bose, E. Le Ru and L. Hardy for 
helpful discussions. D.J. also thanks Mr Vee for helpful distractions. We 
acknowledge the support of the Brazilian agency Conselho Nacional de Desenvolvimento 
Cient\'{\i}fico e Tecnol\'{o}gico (CNPQ), The ORS Award Scheme,
the United Kingdom Engineering and Physical Sciences Research Council, 
The Leverhulme Trust, the European Science Foundation and the European Union.

Note added: After this work was completed, L. Hardy called our attention to his
simultaneous work \cite{Hardy}, in which eq. (1) is also obtained using entirely different methods.

\end{multicols}

\begin{references}
\bibitem{remark3} An `ordered' Schmidt decomposition is one where the
Schmidt coefficients appear in decreasing order. All Schmidt decompositions in 
this paper are of this kind.
\bibitem{remark2}  Here, as in the proof of theorem 1, we consider for
simplicity that all procedures lead to states with the same Schmidt basis.
\bibitem{remark4} For example, it can be shown that the convex roof extension of
$\left\langle E\right\rangle _{\max }$ to mixed states is itself an entanglement monotone
under all LQCC.
\end{references}
\end{document}